# Exploring Knowledge Leakage Risk in Knowledge-Intensive Organisations: behavioural aspects and key controls

*Full Paper*

**Hibah Altukruni**
School of Computing and Information Systems
University of Melbourne
Melbourne, Australia
Email: haltukruni@student.unimelb.edu.au

**Sean Maynard**
School of Computing and Information Systems
University of Melbourne
Melbourne, Australia
Email: seanbm@unimelb.edu.au

**Moneer Alshaikh**
College of Computer Science and Engineering
The University of Jeddah
Saudi Arabia
Email: malshaikh@uj.edu.sa

School of Computing and Information Systems
University of Melbourne
Melbourne, Australia

**Atif Ahmad**
School of Computing and Information Systems
University of Melbourne
Melbourne, Australia
Email: atif@unimelb.edu.au

## Abstract

Knowledge leakage poses a critical risk to the competitive advantage of knowledge-intensive organisations. Although knowledge leakage is a human-centric security issue, little is known about leakage resulting from individual behaviour and the protective strategies and controls that could be effective in mitigating leakage risk. Therefore, this research explores the perspectives of security practitioners on the key factors that influence knowledge leakage risk in the context of knowledge-intensive organisations. We conduct two focus groups to explore these perspectives. The research highlights three types of behavioural controls that mitigate the risk of knowledge leakage: human resource management practices, knowledge security training and awareness practices, and compartmentalisation practices.

# 1  Introduction

Knowledge-intensive Organisations (KIOs) face a critical risk of the erosion of their knowledge assets, which is known as knowledge leakage (KL). Knowledge assets are described as the accumulated intellectual property of KIOs (including knowledgeable workers and their expertise). KIOs create, operate, and implement 'know-how' as a competing product – this is their Intellectual Property (IP). Literature argues that KIOs assign the responsibilities of protecting knowledge assets to security management (Hislop et al. 2018). In security management, the human aspect of a KL threat is more elusive and perplexing than technical aspects (Crossler et al. 2013). This is because individuals may temporarily neglect or ignore knowledge protection controls. Although literature is devoted to various topics concerning knowledge protection and KL (Frishammar et al. 2015), these studies acknowledge limitations in exploring the criticality of KL from the social perspective. Our research aims to identify and present human-generated KL risks and key controls that mitigate them. A key practice-based motivation for this study is to assist security practitioners and scholars to better monitor and manage KL and the potential threats arising from KL. We therefore pose the following research question:

*From security practitioner's perspectives, what key factors affect knowledge leakage risk?*

This research examines the key factors affecting KL behavioural risk. To guide this research, we have explored security practitioners' perspectives on the key facilitators and security controls of KL behavioural risk in the context of KIOs. KIOs are different from traditional organisations, they are "companies where most work can be said to be of an intellectual nature and where well-educated, qualified employees form the major part of the workforce" (Alvesson 2004). For KIOs, knowledge assets are their primary wealth creator and should be considered more valuable than other assets (physical). Further, knowledgeable workers are different from regular employees in that they are authorised to work directly with organisation's confidential knowledge assets (Hislop et al. 2018).

# 2  Background

Discussion of the phenomena of KL occurs in many disciplines, including: knowledge management, innovation management, strategic management, and information security (Durst 2019). Knowledge leakage (KL) is "the deliberate or accidental loss of knowledge to unauthorised personnel within or outside of an organisation boundary" (Ahmad et al. 2014). A substantial body of literature affirms that the erosion of competitive advantage is the largest threat from KL (Jiang et al. 2013). Frishammar et al. (2015) demonstrate that when rivals obtain knowledge about a firm's product, the competitive uniqueness of KIOs would be limited. Ritala et al. (2015) adds that loss of revenue is another consequence of KL. Additionally, the consequence of reputational damage, including doubts about the trustworthiness of a firm that has experienced leaks, can be devastating (Jiang et al. 2013). Building on KL reported in the literature, papers fall into three dimensions: technology (Agudelo-Serna et al. 2016), operational (Frishammar et al. 2015), and human (Durst 2019).

## 2.1  Technology Dimension

Studies highlight that poor security practice is a large source of KL (Durst and Zieba 2019; Hislop et al. 2018). The issues discussed relate mainly to emerging technologies. For example, poor security practices for BYOD are discussed from the perspective of organisational practice and employee behaviour. Allowing employees to BYOD (such as phones and laptops) was the most often reported factor in sensitive KL incidents (Agudelo-Serna et al. 2017). The arising knowledge leakage risk (KLR) seems to be related to poor risk assessment on the part of the organisation, as well as technical issues with the devices themselves (including bugs and vulnerabilities). Agudelo-Serna et al. (2017) argue that KL may also occur through these devices as a result of an organisation failing to sufficiently assess BYOD risks. The improper management of emerging technologies may also lead to KL. Emerging technologies related to KL include cloud computing technologies (Ahmad et al. 2014) and social networking (Christina et al. 2016). Emerging technologies create a number of unexpected KLRs (Hislop et al. 2018). For example, Christina et al. (2016) illustrate that the use of social media poses an internal leakage risk. Another form of leakage occurs with inter-organisational digital collaboration (Jiang et al. 2013). Therefore, if organisations do not securely manage digital communications, employees could leak confidential knowledge through these channels.

## 2.2  Operational Dimension

The operational dimension is widely discussed in the literature in correlation with KLR. It considers securing the firm's internal and external coordination and collaboration activities. Intra-organisational coordination refers to when internal teams from different departments collaborate to perform a task.

Lee et al. (2017) found that limitations in studying internal-to-internal leakage risk facilitated leakage behaviour among employees. Moreover, blurry boundaries in internal coordination also lead to confidential KL (Durst and Zieba 2019; Ritala et al. 2018). If a firm fails to implement adequate security guidelines on knowledge exchange among and within teams, knowledgeable employees may accidently share confidential knowledge with unauthorised employees. Inter-organisational collaboration refers to external activities with other organisations (e.g. outsourcing). Inadequate security practices around these collaborations leads employees to perform intentional or unintentional KL (Durst and Zieba 2019). Jiang et al. (2013) explains this phenomenon: 'If firms exchange their rent-generating knowledge beyond the firm boundary, such as inter-firm strategic alliance, the knowledge is susceptible to expropriation hazards.' Inkpen et al. (2019) suggests that unintentional KL to unauthorised personnel might occur when an employee working for the partner firm accidentally shares business-critical knowledge not meant for exposure. Conversely, Frishammar et al. (2015) demonstrates that a leakage culture may exist in third-party firms as they may deliberately leak KIO innovations to competitors. In this leakage situation, the leakage risk arises due to poor risk assessment practices from the KIOs.

## 2.3　Human Dimension

Although KLR is strongly related to the human dimension, studies investigating the human factors of KL are few. Literature has focused on embedding the human dimension within technology and operations (Ritala et al. 2018). For technology, individuals intentionally or unintentionally share business-critical knowledge through IT artefacts (Christina et al. 2016). The literature in the technology dimension neglects the social aspect of the issue. In the operational dimension, the human attribute is positioned within supply chain or collaborative processes (Lee et al. 2017). Researchers highlight that individuals may leak knowledge inadvertently, or on purpose, during formal or informal occasions (Frishammar et al. 2015; Ritala et al. 2015). Inkpen et al. (2019) illustrates that organisations face a KL issue as employees leave the organisation. This issue is exacerbated when knowledgeable employees join competitors. Further, Nishat Faisal et al. (2007) found a correlation between deliberate KL attempts from employees and incentives offered by competitors. Therefore, the individual employee plays a critical role in how knowledge is leaked. Shabtai et al. (2012) note that with BYOD, employees can carry confidential knowledge with them wherever they go. As a result, some employees may believe that constantly transporting confidential knowledge is a secure behaviour. A recent systematic review (Durst and Zieba 2019) suggests that the KL issue proliferates when organisations do not properly educate their employees about leakage risks associated with mobile devices. Therefore, from the employee's perspective, misuse of portable devices could lead to KL. Agudelo-Serna et al. (2017) found strong evidence that knowledgeable employees could accidentally leak confidential knowledge because they might be unaware of the risks linked to their mobile devices. Therefore, improper training and awareness practices would result in reckless employee behaviour regarding the security of knowledge.

## 2.4　Knowledge Leakage Risk Controls

Our review of the literature on knowledge protection uncovered that organisations should implement several controls to prevent KL behaviour. First, organisations should have an asset classification system that classifies knowledge assets based on their level of sensitivity. Sensitivity classification is an economic decision—that is, the knowledge asset should be classified based on its economic value to the organisation. The impact of that asset being leaked should also be considered in the classification process (Kramer and Bradfield 2010).

Second, the literature highlights the importance of human resource practices in preventing KLR. Olander et al. (2015) concluded that HR departments that communicate protection responsibilities to employees tend to obtain higher employee commitment and loyalty towards organisation assets. Moreover, they highlight the importance of ongoing monitoring activities that profile employees to understanding the risk associated with them. Shaw and Stock (2011) report that organisations are dependent upon their Intellectual Property (IP) should institute prior hiring practices such as background checks and interviews.

Third, creating a security culture and making employees aware has been also widely discussed in the literature. Training and awareness practices are aimed to make employees more vigilant about security risks (Alshaikh et al. 2018). Yalabik et al. (2017) suggest that training programs might change the employees' negative traits as they relate to the organisation's assets and increase their commitment to the organisation. However, Yalabik et al. (2017) suggests that training and awareness practices would increase the employees' consciousness of social engineering attempts from authorised individuals.

Lastly, restricted access of knowledge assets has been discussed in the KL literature in terms of compartmentalisation (Ahmad et al. 2014; Baldwin and Henkel 2015). Compartmentalisation in information security is described as a restriction on accessing and communicating an organisation's confidential assets between different groups (Elliott et al. 2019). Compartmentalisation can be implemented physically, where organisations have different offices for knowledgeable employees—that would increase protection of an organisation's valuable resources and limited KL.

## 3　Research Methodology

This research was undertaken with an exploratory, qualitative design to investigate security practitioners' views of the key factors of KL in KIOs. The design allows for a better understanding of a particular issue (Neuman 2014); therefore, it has been used in information system studies to seek a better understanding of specific information systems cases (Shedden et al. 2016) and, more specifically, behavioural information security research (Chatterjee et al. 2015; Crossler et al. 2013). We first propose a descriptive KL framework (see Section 4). Second, we conduct 2 focus groups using the generated scenarios in our KL framework. We invited 10 participants to undertake the research across two focus groups of 5 each, however for the second focus group only two participants were able to attend (see Table 1 for participants). Using the focus group technique allowed participants to exchange views and further explain their ideas and experiences (Creswell 2018). After completing the group discussions, data saturation was achieved, as no new information from the analysed data emerged (Ness 2015). We used thematic analysis to analyse the focus group data as it provides the accessibility and flexibility to reassess and analyse the collected data.

| ID | Experience (Years) | Industry | Role |
|---|---|---|---|
| P1 | 17+ | ICT | Information Technology Manager |
| P2 | 13+ | Education | Security Analyst |
| P3 | 10+ | Software services | Senior Security Manager |
| P4 | 7+ | Consulting | Information Security Consultant |
| P5 | 26+ | Consulting | Information Security Consultant |
| P6 | 20+ | Government and Research | Chief Information Security Officer |
| P7 | 28+ | ICT | Security Manager |

*Table 1 Participants' characteristics.*

## 4　Research Design

As KLR is a human-centred phenomenon, exploring KLR through a behavioural information security lens is required (Hislop et al. 2018). Crossler et al. (2013) describe behavioural information security research as a "subfield of the broader InfoSec field that focuses on the behaviours of individuals which [is] related to information and information systems assets". Crossler et al. (2013) recommends utilising reported, instead of self-reported, behavioural scenarios to better reflect the objectivity of building the data collection questionnaire and increasing data reliability, as intentions do not always lead to action. Therefore, the authors have opted to extensively explore reported KL incidents in academic and professional literature. KL poses a threat to the organisation's confidential knowledge assets. To better understand this threat, security researchers have argued that modelling is a good tool for providing a bigger picture of security issues (Shostack 2014). Therefore, as this research's aim is to explore KL behavioural risk, modelling this risk assists in identifying examples of knowledge disclosure. One of the key strategies in structuring the threat modelling is focusing on attackers (Shostack 2014). As KL behavioural risk is a human-centric issue, an attacker-driven strategy is adopted. This approach specifies possible approaches used by people who might intentionally or unintentionally attack the organisation's assets. (Shostack 2014) argue that an attacker-driven approach is helpful when the study aims to explain who might attack the organisation assets and humanises the risk by adding a human threat agent. As studies investigating the key factors of KL behavioural risk are underdeveloped, the data leakage scenario classification parameters suggested by Shabtai et al. (2012) guide the development of our KL scenarios. Accorsi et al. (2015) illustrate that the data leakage classification assists identifying potential covert leakage channels; therefore, participants would be exposed to KL scenarios that they might not previously have experienced. To better validate our generated KL scenarios, the researchers iteratively reviewed and synthesised KL incidents reported in the literature against the developed framework.

### 4.1　KL Behavioural Framework

Conceptualising confidential KLR is based on five parameters (adapted from Shabtai et al. (2012)): where the KL occurs, who caused the KL (human threat agent), how access to the knowledge was gained, the nature of the leakage, and how the knowledge leaked.

#### 4.1.1 Where did knowledge leakage occur?

The first parameter refers to the possible locations of KL, which can be divided into three areas. First, inside the organisation, meaning the KL incident occurs within the organisation's physical boundaries (Hislop et al. 2018), such as two teams from different departments working collaboratively. Second, outside the organisation, meaning the KL incident occurs outside the organisation's physical boundaries (Frishammar et al. 2015), such as confidential new product designs were stolen from an employee's computer outside the organisation. In the third method, the KL occurs at a third-party's physical location (Ritala et al. 2018).

#### 4.1.2 Who caused knowledge leakage?

The second parameter highlights the possible human threat agents who would leak knowledge, which follows the KL locations: an insider, an outsider, or an external insider. An insider is an internal trusted source resident within the organisation, such as a knowledge worker or other regular employee (Hislop et al. 2018). Outsiders mean that the KL or stealing occurs from an external source, such as competitors attempting to steal knowledge assets (Inkpen et al. 2019). External insiders mean that the KL source is a trusted partner who has worked in the organisation but is currently working elsewhere, such as a contractor (consultancy) (Ritala et al. 2018).

#### 4.1.3 How was access to knowledge gained?

The third parameter explains the access privilege of the KL source. Access privilege to knowledge assets can be placed into two categories: authorised or unauthorised (Hislop et al. 2018). In KL scenarios, authorised refers to trusted individuals who obtain a legitimate access to knowledge assets —either insiders such as knowledge workers or external insiders such as suppliers (Shabtai et al. 2012). Unauthorised individuals would be insiders who do not have legal access to knowledge assets, as their daily work does not require accessing confidential knowledge such as new interns or employees from different departments (Lee et al. 2017) or competitors (Frishammar et al. 2015).

#### 4.1.4 The nature of leakage

The fourth parameter indicates the intention of the person leaking knowledge – whether it is deliberate or accidental (Agudelo-Serna et al. 2015). Accidental means knowledge was unwittingly leaked as a result of performing an organisation process or participating in an informal event (Bulgurcu et al. 2010). Deliberate means that knowledge was leaked on purpose with malicious intent, ignoring organisational security and knowing the confidentiality of the knowledge and the occupational hazards of the KL incident (Nishat Faisal et al. 2007).

#### 4.1.5 The medium of knowledge leakage

The last parameter highlights the KL medium: physical, digital, or conversational. Distinguishing between the mediums might be useful, as the discussion around knowledge protection would be different. Physical KL occurs through physical means (documents, diagrams, or handwritten notes) (Durst and Zieba 2019); digital leaks occur through emails, social media, BYOD, and intranet (Ilvonen et al. 2018). Conversational KL means verbal leakage during occasions like professional conferences, Q&A sessions (Ritala et al. 2018); or informal occasions such as casual events (Olander et al. 2015).

#### 4.1.6 Knowledge leakage behaviour scenarios

Table 2 shows our framework that combines each of the parameters of KL. We have also included representative references for the KL parameter combinations. Eighteen Generic KL Scenarios were developed based on the KL parameters discussed (see Appendix A for sample scenarios).

## 5 Findings

Two main themes were identified in our analysis: interpersonal enabling factors and organisational practices around KL mitigation. Sub themes included: KL behaviour and employees' personality traits, poor knowledge classification practices, and poor knowledge security management practices.

### 5.1 Individual-level enabling factors

Participants discussed the individual-level factors that contributed to KL behaviour, which highlighted two sub-themes. The first sub-theme is KL behaviour. Participants suggested that some employees' KL behaviours are difficult to address with security controls. As P1 states: *'You can't label word of mouth ... for now, we have no formal or technological controls about what people can say or not say'.* Employee mobility is also an issue with P1 suggesting *'…leaking is when an employee of organisation A that has knowledge of how that organisation operates moves to organisation B'*. Furthermore, P4 suggests that this is even harder from a social engineering aspect, '*The Threat Ops… went after competitors across the globe to collect human interventions'.* (P4).

The second sub-theme is employees' motivation. During the discussion, participants stated that an employee's intrinsic/extrinsic motivation might influence his/her tendencies to engage in unauthorised disclosure of confidential knowledge through KL risk behaviours. This theme was elaborated as: a sense of entitlement with P3 stating: '*… to share that knowledge is he wants to feel empowered or he wants to feel important'*. Additionally, P6 discussed the employees' ambition to protect or advance themselves: '*… Maybe for their own benefit, as you can say to get more business—to share that information with a competitor'.* Finally, P2 comments on motivation from a competitor nature perspective: '*Maybe the person is competitive in nature and he wants to boast about his projects'.*

| Parameters of KL ||||| Representative References | Scenario |
|---|---|---|---|---|---|---|
| Where Leakage Occurs | Cause of Leakage | Authorised/ Unauthorised Access | Leakage Nature | Leakage Medium | | |
| Inside | Insider | Authorised | Accidental | Physical | (Hislop et al. 2018; Ritala et al. 2015) | A |
| Inside | Insider | Authorised | Accidental | Digital | | B |
| Inside | Insider | Authorised | Accidental | Conversation | | C |
| Inside | Insider | Authorised | Deliberate | Physical | (Hislop et al. 2018; Ritala et al. 2015) | D |
| Inside | Insider | Authorised | Deliberate | Digital | | E |
| Inside | Insider | Authorised | Deliberate | Conversation | | F |
| Inside | Insider | Unauthorised | Deliberate | Physical | (Lee et al. 2017) | G |
| Inside | Insider | Unauthorised | Deliberate | Digital | | H |
| Inside | Insider | Unauthorised | Deliberate | Conversation | | I |
| Outside | Outsider | Unauthorised | Deliberate | Physical | (Frishammar et al. 2015) | J |
| Outside | Outsider | Unauthorised | Deliberate | Digital | | K |
| Outside | Outsider | Unauthorised | Deliberate | Conversation | | L |
| 3rd Party | Ext Insider | Authorised | Accidental | Physical | (Jiang et al. 2013; Ritala et al. 2015) | M |
| 3rd Party | Ext Insider | Authorised | Accidental | Digital | | N |
| 3rd Party | Ext Insider | Authorised | Accidental | Conversation | | O |
| 3rd Party | Ext Insider | Authorised | Deliberate | Physical | (Frishammar et al. 2015; Shabtai et al. 2012) | P |
| 3rd Party | Ext Insider | Authorised | Deliberate | Digital | | Q |
| 3rd Party | Ext Insider | Authorised | Deliberate | Conversation | | R |

*Table 2: Knowledge leakage behavioural framework showing scenarios.*

## 5.2 Organisational practices around knowledge leakage mitigation

The second main theme identified was organisation practices around KL mitigation and is characterised around poor knowledge sensitivity classification and poor knowledge security practices. Participants discussed poor knowledge sensitivity classification in some detail. P6 said *'Potentially, these organisations don't have classification of the information and they're enabling people to send attachments outside the organisation.' (Q2).* P1 indicated that this might just be poor labelling practices: *'…But for that classification, do they have labelling? If it's a confidential document, was it appropriately labelled?',* however P3 stated that this might go further and be an improper balance of classification between what needs to be shared and what needs to be confidential '*I think that part of the problem is organisations need to know what is IP … Too much of a lockdown will curtail collaboration between organisations. Too much of openness, scenarios will happen like this'.*

Poor knowledge security management practices refers to the security practices applied within the organisation. Participants revealed consensus on the key role that organisation security practices play in facilitating or controlling KL behaviour. Participants discussed the incompleteness of some knowledge security management practices influencing KL behaviour. These inconsistencies included poor internal risk assessment practices: *'Whenever there's a major change within an organisation where people will be demoted or people will be let go, the organisation should be more vigilant in terms of who should have access to what. And who is actually accessing what' (P2);* poor compartmentalisation practices: *'limit the CFO's access to information that relates to the financing, not* *to the technical documentation, because regardless of how senior they are, they don't need access to designs and technologies' (P7).* In terms of controlling KL behaviour, participants agreed on the criticality of three security controls: human resource management practices: *'Which is what it's about so much of the time, isn't it? It's about people: getting the right person in the first place' (P6);* knowledge security training and awareness practices: *'organisations will be going to have a good HR… you can build it [KL scenario] into your security training and awareness scheme … can train on different things [KL scenarios]" (P7);* and compartmentalisation practices: *'But they're [colleagues] working at two separate projects. He shouldn't have access to the other project's documentation. There should be proper segregation' (P5).*

# 6　Discussion

In the findings, we identified components related to interpersonal enabling factors and organisational practices around KL mitigation. In this discussion we focus on three key security controls that had superior impacts in inhibiting the identified key factors of KL.

## 6.1　Human Resource Management Practices

Human resource management (HRM) practices refer to how human resources work to mitigate KL behavioural risk. We found that HRM practices inhibit some of the identified key factors: KL behaviour, employees' personality traits, and poor knowledge management practices. KL occurs through accidental sharing behaviours (Ritala et al. 2015). Our findings extend this and demonstrate that HRM is a key control for mitigating accidental leakage. First, HRM practices help to increase employees' sense of responsibility; reminding them of their shared roles as protectors of knowledge assets. This confirms Hislop et al.'s (2018's) work who states that increasing an employee's sense of responsibility increases their level of confidential knowledge handling care. Extending this, a significant sense of responsibility is also reflected by the employees decreased susceptibility to social engineering. Further, our findings suggest that good HRM practices improve knowledge protection, as they contribute to employees' commitment to that knowledge. Inkpen et al. (2019) argue that managing KL behavioural risk via an employee mobility channel doesn't help to protect organisational knowledge. However, our participants confirm employee mobility as an enabler for KL arguing that controlling KL behaviour via managing employees' mobility was ideal. In this study, participants demonstrate that HRM security practice can mitigate KL via managing employees' mobility behaviour.

Employees' motivations have been largely neglected in KL literature. Our participants argue that a key HRM security practice is proper hiring practice, which would exclude people whose attributes make them seem more vulnerable than others. HRM hiring activities include pre-employment screening, background checks, psychological screening, and interviews. For instance, hiring practices could assess intrinsic / extrinsic motivations (using self-determination theory (Deci and Ryan 2012; Ryan and Deci 2017) of employees to assess how likely an employee needs to have these motivations satisfied. Also, organisations can develop interventions to reduce the impact on extrinsic motivations to reduce their impact (Brown and Kasser 2005). Shaw and Stock (2011) argued that organisations that perform pre-employment background checks would be in a better position to mitigate insiders' deliberate breaching behaviour. This study affirms that KIOs should be more vigilant in selecting their employees, whether regular employees or third parties, especially when KIOs' activities involve IP. According to Lee et al. (2017), internal monitoring practices create a protected internal atmosphere. Our findings suggested that continued monitoring would improve the overall work climate that deters KL behaviour.

Participants reported that embedding HRM practices within internal risk assessment practices is beneficial, especially when organisational changes occur. Shaw and Stock (2011) reported that when organisational changes are improperly introduced to the employees—meaning without assessing the internal risks that would arise from these changes—KL behaviour might occur. The findings expanded on that claim by emphasising HRM security practices' role in monitoring employees' behaviour and risk profiling them. HRM practices with effective reinforcement include communicating the value proposition of the security controls to the employees. According to Bulgurcu et al. (2010), employees' understanding of the value of security controls increases their compliance with these controls. We saw a similar outcome, adding that developing employees' understanding of the benefits and importance of complying with knowledge protection controls influences their compliant behaviour, thus mitigating KL behavioural risk.

## 6.2　Knowledge Security Training and Awareness Practices

Knowledge security training and awareness practices are key factors for mitigating the KL behavioural risk. Alshaikh et al. (2018) argued that training and awareness controls underpin the organisation's security controls. Participants confirmed this, adding that improper knowledge security training and awareness practices undermine KL controls that an organisation applies. Good knowledge security training and awareness practices affect employee KL behaviour. Effective awareness campaigns focus on increasing awareness of employees of the consequences of their risky behaviour and how to handle confidential knowledge. Kraemer et al. (2009) reports that many employees' mistakes were results of inadequate awareness campaigns. The findings of this study illustrated the effectiveness of situational training and awareness programs, where security practitioners design them to raise employees' understanding of KL or social engineering attempts.

Training and awareness practices might lower KL behavioural risk due to employee personality traits. Elliott et al. (2019) highlight personality traits that some employees bring to the workplace, making them vulnerable. They argue that employees who are angry at the organisation would pose a threat to the training program's aims. Our participants suggested training practices would improve employees' commitment and loyalty to the organisation. Yalabik et al. (2017) recommended that effective training activities, such as developing employees' personal skills and knowledge, could result in greater loyalty to the organisation. Therefore, our results suggest that when KIOs train and empower their employees, employee commitment to the organisation increases.

Poor understanding or execution of classification around knowledge in the organisation is reflected in employee KL behaviours. Situational training programs about the knowledge sensitivity classification systems maintained by the organisation would lower KL behavioural risk. Kraemer et al. (2009) argues that the lack of employee training about the organisation's security might make them more vulnerable to information risks. The study participants questioned whether employees were trained about the knowledge classification system, arguing that if KIOs proposed situational training programs that aimed to educate the employees, many of the presented scenarios would not occur.

Training and awareness practices are connected to the deficiency in knowledge security management and poor knowledge risk management practices. Our participants illustrate that situational training on different possible scenarios has a positive impact on the employees' KL behaviour. Shaw et al. (2009) argues that sufficient training activities would encourage employees to be more mindful of their own behaviour and their colleagues' behaviour. Our study's participants highlight training and awareness practices that aim to make employees cautious and more alert, reducing accidental KL behaviour. Participants complained that inadequate training occurs on incident reporting practices, possibly due to employees being discouraged from reporting these incidents. Training and awareness programs that encourage employees to report KL events would aid mitigation of KL behavioural risk: first, by helping security practitioners identify potential KL risks and second, through security practitioners updating training programs to mitigate new KL risks. Hence, reducing accidental KL behaviour.

## 6.3　Compartmentalisation Practices

Our findings question the effectiveness of compartmentalisation as a strategy when security aims to protect knowledge from behavioural KL. Our participants argue that as knowledge is considered learning, meaning employees permanently absorb that knowledge, compartmentalisation has limited security value for mitigating KL behaviour. They argue that in the context of KIOs, organisational change is constant and transformations, such as those where the organisational hierarchy changes, can result in sub optimal security arrangements. For example, in an organisational restructure, employees can change roles or have different responsibilities, however the knowledge from their past is not forgotten. Also, in changing responsibilities they may gain access to more knowledge, building upon knowledge from previous positions. What they know before does not affect what they know later; this is not an issue. However, when their role changes and what the employees know from previous positions becomes a liability, how can organisations compartmentalise individuals with all their background knowledge?

Compartmentalisation practices can be used to mitigate accidental KL incidents. Baldwin and Henkel (2015) assert that the segmentation of knowledgeable employees from others can mitigate confidential knowledge from unauthorised access. Our participants clearly recognise that segmenting employees who work on confidential knowledge from others eliminates some accidental KL scenarios. However, this may be an undesired practice as it might negatively affect the secrecy of the overall confidential knowledge. Instead, organisations should compartmentalise at the knowledge level, with each knowledgeable employee being assigned a specific subset of the knowledge asset. So, even if the knowledgeable employee accidentally or deliberately leaks that subset knowledge, KL will have a lesser impact. This specific partitioning is a better solution where it is difficult for an individual to have the whole picture of the organisation's confidential knowledge.

A knowledge sensitivity classification system is an organisation's understanding of what it has and how sensitive it is. Elliott et al. (2019) reported secrecy to be an essential element of knowledge sensitivity classification and that balancing secrecy measures and functionality is a necessity for controlling KL. Similarly, our participants highlight classification practices in which too much compartmentalisation might lead employees to bend the rules and accidentally leak knowledge. Participants highlight that locking a cabinet every time an employee is away, for example, is not practical; as a result, the employees might think that taking the confidential knowledge with them is safer and more convenient.

Our participants pointed out a critical risk, which is that the majority of the reported KL scenarios are caused by authorised sources. As the threat agents need to know the knowledge to complete their work, these cases are problematic, especially when collaborating with third parties. Previous studies describe the third-party dilemma as the 'paradox of openness' (Elliott et al. 2019). Our participants shared the same views and added the risk of knowledgeable insider employees, as they also have a legitimate access to the knowledge. The security practitioners argued that the current practices of compartmentalisation would have limited value at this stage.

## 7  Contribution and Conclusion

In this paper we identify the key factors that affect KL in organisations. These factors are evident in an organisation's human resource management, knowledge security and awareness training, and compartmentalisation practices. Altogether, these findings demonstrate that addressing the human dimension within KL risk research is an essential practice for mitigating KL incidents.

This research contributes first to methodology as we construct an instrument for information collection (Thomas 2003). To generate the KL scenarios this study developed a KL behavioural risk framework. According to (Crossler et al. 2013), using scenarios in studying behavioural information security studies is preferable, as it might expose study participants to scenarios that may be new to them. Using scenario-based questions, the study's participants reported encountering new behavioural KL incidents. Moreover, our findings differentiate between the KL mediums, due to the use of the scenarios. Our participants highlighted that physical and conversational means are the most problematic to mitigate, whereas there are plenty of digital controls to mitigate KL. Second, from a research contribution perspective we contribute in two ways. 1) we identify interpersonal KL enabling factors, which are usually neglected in KL studies. Our study identifies human aspects linked to KL risk and complements the missing view of understanding KLR. 2) this research disagrees with previous information security literature regarding the extent to which compartmentalisation would be helpful in mitigating KL. This is because knowledge as a concept involves learning and interpretation; current compartmentalisation practices would not be a potent control if the KL source was authorised and has legitimate access to confidential knowledge. In fact, many reported KL scenarios are from authorised individuals. Therefore, this study provides an alternative approach of compartmentalisation. We also make contributions to practice. The innovation literature claims there are few KL controls for an organisation to apply and there are uncertain benefits from them (Inkpen et al. 2019). Our research suggests that security practitioners invest in practices to reduce KL: HRM practices, knowledge security training and awareness practices, and compartmentalisation.

Our research suggests that organisation practices affect KL behaviour; therefore, future research should investigate organisational characteristics in the developed framework. The investigation of alternative compartmentalisation strategies to explore the extent of how to mitigate KL is useful. Further, examining the physical and conversational mediums in more detail, as our findings indicate that KL from physical mediums is difficult to track and easier to be stolen or lost, and conversational mediums are difficult to control, as they involve personal judgment and awareness.

## 8  References

Accorsi, R., Lehmann, A., and Lohmann, N. 2015. "Information Leak Detection in Business Process Models: Theory, Application, and Tool Support," *Information Systems* (47), pp 244-257.
Agudelo-Serna, C.A., Bosua, R., Ahmad, A., and Maynard, S.B. 2015. "Understanding Knowledge Leakage & Byod ( Bring Your Own Device ): A Mobile Worker Perspective," in: *The 26th Australasian Conference on Information Systems*. pp. 1-13.
Agudelo-Serna, C.A., Bosua, R., Ahmad, A., and Maynard, S.B. 2016. "Mitigating Knowledge Leakage Risk in Organizations through Mobile Devices: A Contextual Approach," *27th Australasian*

*Conference on Information Systems*, Wollongong, Australia: University of Wollongong, p. 12 pages.

Agudelo-Serna, C.A., Bosua, R., Ahmad, A., and Maynard, S.B. 2017. "Strategies to Mitigate Knowledge Leakage Risk Caused by the Use of Mobile Devices: A Preliminary Study," *ICIS*, Seoul, South Korea, p. 19.

Ahmad, A., Bosua, R., and Scheepers, R. 2014. "Protecting Organizational Competitive Advantage: A Knowledge Leakage Perspective," *Computers and Security* (42), pp 27-39.

Alshaikh, M., Maynard, S.B., Ahmad, A., and Chang, S. 2018. "An Exploratory Study of Current Information Security Training and Awareness Practices in Organizations," *Proceedings of the 51st Hawaii International Conference on System Sciences*, Hawaii, p. 10.

Alvesson, M. 2004. *Knowledge Work and Knowledge-Intensive Firms*. OUP Oxford.

Baldwin, C.Y., and Henkel, J. 2015. "Modularity and Intellectual Property Protection," *Strategic management journal* (36:11), pp 1637-1655.

Brown, K.W., and Kasser, T. 2005. "Are Psychological and Ecological Well-Being Compatible? The Role of Values, Mindfulness, and Lifestyle," *Social Indicators Research* (74:2), pp 349-368.

Bulgurcu, B., Cavusoglu, H., and Benbasat, I. 2010. "Information Security Policy Compliance: An Empirical Study of Rationality-Based Beliefs and Information Security Awareness," *MIS Quarterly* (34:3), pp 523-527.

Chatterjee, S., Sarker, S., and Valacich, J.S. 2015. "The Behavioral Roots of Information Systems Security: Exploring Key Factors Related to Unethical It Use," *Journal of Management Information Systems* (31:4), pp 49-87.

Christina, S., Stefan, T., and Markus, M. 2016. "Protecting Knowledge in the Financial Sector: An Analysis of Knowledge Risks Arising from Social Media," *2016 49th Hawaii International Conference on System Sciences (HICSS)*: IEEE, pp. 4031-4040.

Creswell, J.W. 2018. *Qualitative Inquiry and Research Design: Choosing among Five Traditions*, (fourth ed.). Thousand Oaks, CA: Sage Publications.

Crossler, R.E., Johnston, A.C., Lowry, P.B., Hu, Q., Warkentin, M., and Baskerville, R. 2013. "Future Directions for Behavioral Information Security Research.," *Computer & Security* (23), pp 90-101.

Deci, E.L., and Ryan, R.M. 2012. "Motivation, Personality, and Development within Embedded Social Contexts: An Overview of Self-Determination Theory," *The Oxford handbook of human motivation*), pp 85-107.

Durst, S. 2019. "How Far Have We Come with the Study of Knowledge Risks?," *VINE Journal of Information and Knowledge Management Systems* (49:1), pp 21-34.

Durst, S., and Zieba, M. 2019. "Mapping Knowledge Risks: Towards a Better Understanding of Knowledge Management," *Knowledge Management Research & Practice* (17:1), pp 1-13.

Elliott, K., Patacconi, A., Swierzbinski, J., and Williams, J. 2019. "Knowledge Protection in Firms: A Conceptual Framework and Evidence from Hp Labs," *European Management Review* (16:1), pp 179-193.

Frishammar, J., Ericsson, K., and Patel, P.C. 2015. "The Dark Side of Knowledge Transfer: Exploring Knowledge Leakage in Joint R&D Projects," *Technovation* (41-42), 7/1/July-August 2015, pp 75-88.

Hislop, D., Bosua, R., and Helms, R. 2018. *Knowledge Management in Organizations: A Critical Introduction*, (fourth ed.).

Ilvonen, I., Thalmann, S., Manhart, M., and Sillaber, C. 2018. "Reconciling Digital Transformation and Knowledge Protection: A Research Agenda," *Knowledge Management Research & Practice* (16:2), pp 235-244.

Inkpen, A., Minbaeva, D., and Tsang, E.W. 2019. "Unintentional, Unavoidable, and Beneficial Knowledge Leakage from the Multinational Enterprise," *Journal of International Business Studies* (50:2), pp 250-260.

Jiang, X., Li, M., Gao, S., Bao, Y., and Jiang, F. 2013. "Managing Knowledge Leakage in Strategic Alliances: The Effects of Trust and Formal Contracts," *Industrial Marketing Management* (42), pp 983-991.

Kraemer, S., Carayon, P., and Clem, J. 2009. "Human and Organisational Factors in Computer and Information Security: Pathways to Vulnerabilities," *Computers & Security* (28), pp 509-520.

Kramer, S., and Bradfield, J.C. 2010. "A General Definition of Malware," *Journal in Computer Virology*:2), p 105.

Lee, J., Min, J., and Lee, H. 2017. "Setting a Knowledge Boundary across Teams: Knowledge Protection Regulation for Inter-Team Coordination and Team Performance," *Journal of Knowledge Management* (21:2), pp 254-274.

Ness, L.R. 2015. "Are We There Yet? Data Saturation in Qualitative Research,").

Neuman, W.L. 2014. *Social Research Methods: Qualitative and Quantitative Approaches*, (Seventh ed.). London: Pearson Education Ltd.

Nishat Faisal, M., Banwet, D.K., and Shankar, R. 2007. "Information Risks Management in Supply Chains: An Assessment and Mitigation Framework," *Journal of Enterprise Information Management* (20:6), pp 677-699.

Olander, H., Hurmelinna-Laukkanen, P., and Heilmann, P. 2015. "Human Resources - Strength and Weakness in Protection of Intellectual Capital," *Journal of Intellectual Capital* (16:4), pp 742-762.

Ritala, P., Husted, K., Olander, H., and Michailova, S. 2018. "External Knowledge Sharing and Radical Innovation: The Downsides of Uncontrolled Openness," *Journal of Knowledge Management* (22:5), 2018/06/11, pp 1104-1123.

Ritala, P., Olander, H., Michailova, S., and Husted, K. 2015. "Knowledge Sharing, Knowledge Leaking and Relative Innovation Performance: An Empirical Study," *Technovation* (35), pp 22-31.

Ryan, R.M., and Deci, E.L. 2017. *Self-Determination Theory: Basic Psychological Needs in Motivation, Development, and Wellness*. Guilford Publications.

Shabtai, A., Elovici, Y., and Rokach, L. 2012. "A Survey of Data Leakage Detection and Prevention Solutions,"), p 92.

Shaw, E.D., and Stock, H.V. 2011. "Behavioral Risk Indicators of Malicious Insider Theft of Intellectual Property: Misreading the Writing on the Wall," *White Paper, Symantec, Mountain View, CA* (2011).

Shaw, R.S., Chen, C.C., Harris, A.L., and Huang, H.-J. 2009. "The Impact of Information Richness on Information Security Awareness Training Effectiveness," *Computers & Education* (52:1), pp 92-100.

Shedden, P., Ahmad, A., Smith, W., Tscherning, H., and Scheepers, R. 2016. "Asset Identification in Information Security Risk Assessment: A Business Practice Approach," *Communications of the Association for Information Systems*).

Shostack, A. 2014. *Threat Modeling: Designing for Security*. John Wiley & Sons.

Thomas, R.M. 2003. "Contributing to Research Methodology Theses and Dissertations," in: *Blending Qualitative and Quantitative Research Methods in Theses and Dissertations*. Corwin Press, pp. 171-198.

Yalabik, Z.Y., Swart, J., Kinnie, N., and Van Rossenberg, Y. 2017. "Multiple Foci of Commitment and Intention to Quit in Knowledge-Intensive Organizations (Kios): What Makes Professionals Leave?," *The International Journal of Human Resource Management* (28:2), pp 417-447.

## Appendix A – Examples of Scenarios

Scenario C: *Sam works as a knowledge manager at Tech Company. At a conference, he talks to a group of managers from different companies. Two of the managers talk about the outcomes of a major project that they collaboratively worked on. Sam feels that he is being left out so starts talking about a new project his company is working on. In doing so, Sam inadvertently divulges some secret details while bragging.*

Scenario F: *Tina is the CFO at Innovative Technologies. The company, as a result of a governance audit, is restructured which results in demotion for Tina to a non-executive position. As a result, Tina, whilst complaining to a colleague in a rival firm, deliberately discloses trade secrets relating to the designs of her firm's latest innovation.*

Scenario R: *During a visit to Organisation A, Jeff, a consultant, learns about their intentions regarding critical upcoming investments. Later whilst on another consulting assignment, Jeff discusses these intentions with a competitor to Organisation A.*